\documentclass[%
 aip,
 amsmath,amssymb,
 reprint,%
]{revtex4-1}

\usepackage{graphicx}
\usepackage{dcolumn}
\usepackage{bm}

\usepackage[utf8]{inputenc}
\usepackage[T1]{fontenc}
\usepackage{mathptmx}

\begin{document}

\preprint{AIP/123-QED}

\title{Silicon Substitution in Nanotubes and Graphene via Intermittent Vacancies}

\author{Heena Inani}
\author{Kimmo Mustonen}
\email{kimmo.mustonen@univie.ac.at.}
\author{Alexander Markevich}
\affiliation{ 
Faculty of Physics, University of Vienna, Boltzmanngasse 5, 1090 Vienna, Austria}%
\author{Er-Xiong Ding}
\affiliation{ 
Aalto University School of Science, Department of Applied Physics, P.O. Box 15100, FI-00076 Aalto, Finland}%
\author{Mukesh Tripathi}
\affiliation{ 
Faculty of Physics, University of Vienna, Boltzmanngasse 5, 1090 Vienna, Austria}%
\author{Aqeel Hussain}
\affiliation{ 
Aalto University School of Science, Department of Applied Physics, P.O. Box 15100, FI-00076 Aalto, Finland}%
\author{Clemens Mangler}
\affiliation{ 
Faculty of Physics, University of Vienna, Boltzmanngasse 5, 1090 Vienna, Austria}%
\author{Esko I. Kauppinen}
\affiliation{ 
Aalto University School of Science, Department of Applied Physics, P.O. Box 15100, FI-00076 Aalto, Finland}%
\author{Toma Susi}
\author{Jani Kotakoski}
\email{jani.kotakoski@univie.ac.at.}
\affiliation{ 
Faculty of Physics, University of Vienna, Boltzmanngasse 5, 1090 Vienna, Austria}%
\date{\today}

\begin{abstract}
The properties of single-walled carbon nanotubes (SWCNTs) and graphene can be modified by the presence of covalently bound impurities. Although this can be achieved by introducing chemical additives during synthesis, that often hinders growth and leads to limited crystallite size and quality. Here, through the simultaneous formation of vacancies with low-energy argon plasma and the thermal activation of adatom diffusion by laser irradiation, silicon impurities are incorporated into the lattice of both materials. After an exposure of $\sim$1~ion/nm$^{2}$, we find Si substitution densities of 0.15~nm$^{-2}$ in graphene and 0.05~nm$^{-2}$ in nanotubes, as revealed by atomically resolved scanning transmission electron microscopy. In good agreement with predictions of Ar irradiation effects in SWCNTs, we find Si incorporated in both mono- and divacancies, with $\sim$2/3 being of the first type. Controlled inclusion of impurities in the quasi-1D and 2D carbon lattices may prove useful for applications such as gas sensing, and a similar approach might also be used to substitute other elements with migration barriers lower than that of carbon.
\end{abstract}

\maketitle

Graphene~\cite{geim2010rise} and single-walled carbon nanotubes~\cite{iijima1993single} (\mbox{SWCNTs}) are among the most studied materials of the last two decades. Due to the confinement in either one or two dimensions and the fact that they consist exclusively of surface atoms, their properties such as electronic transport and chemical reactivity are highly sensitive to any structural perturbations.~\cite{ novoselov2004electric, baierle2001electronic} Covalent incorporation of foreign atoms within their lattice has thus been proposed as a viable route to engineer their properties.~\cite{Ewels05JNN,Terrones08,Ayala10RMP}

Atomic-scale observations in graphene have shown nitrogen~\cite{bangert2013ion} (N), boron~\cite{bangert2013ion} (B), phosphorus~\cite{susi2017single} (P), silicon~\cite{ramasse2013probing} (Si) and germanium~\cite{tripathi2018implanting} (Ge) either as naturally occurring or substitutionally implanted covalent impurities. In SWCNTs, the presence of N,~\cite{Keskar05CPL,Susi11CM} B,~\cite{Ayala10APL} P~\cite{campos2010chemical,Ruiz-Soria15C} and Si~\cite{campos2010chemical} has been spectroscopically detected in chemically synthesized samples, but no direct transmission electron microscopy (TEM) evidence of their incorporation in the tube walls has been shown. Only for N has such conclusive evidence been presented by Arenal \textit{et al.}~\cite{arenal2014atomic} Si impurities have attracted particular recent attention due to the possibility of manipulating them with the focused electron beam.~\cite{susi_silicon_2014,dyck2017placing,tripathi_electron-beam_2018} As a route complementary to chemical synthesis, Dyck \textit{et al.} recently used a 100~kV scanning transmission electron microscopy (STEM) probe to create vacancies in graphene, which they then managed to fill with Si by irradiating the surrounding amorphous contamination.~\cite{dyck2017placing}

\begin{figure}
\includegraphics[width=8.6cm,keepaspectratio]{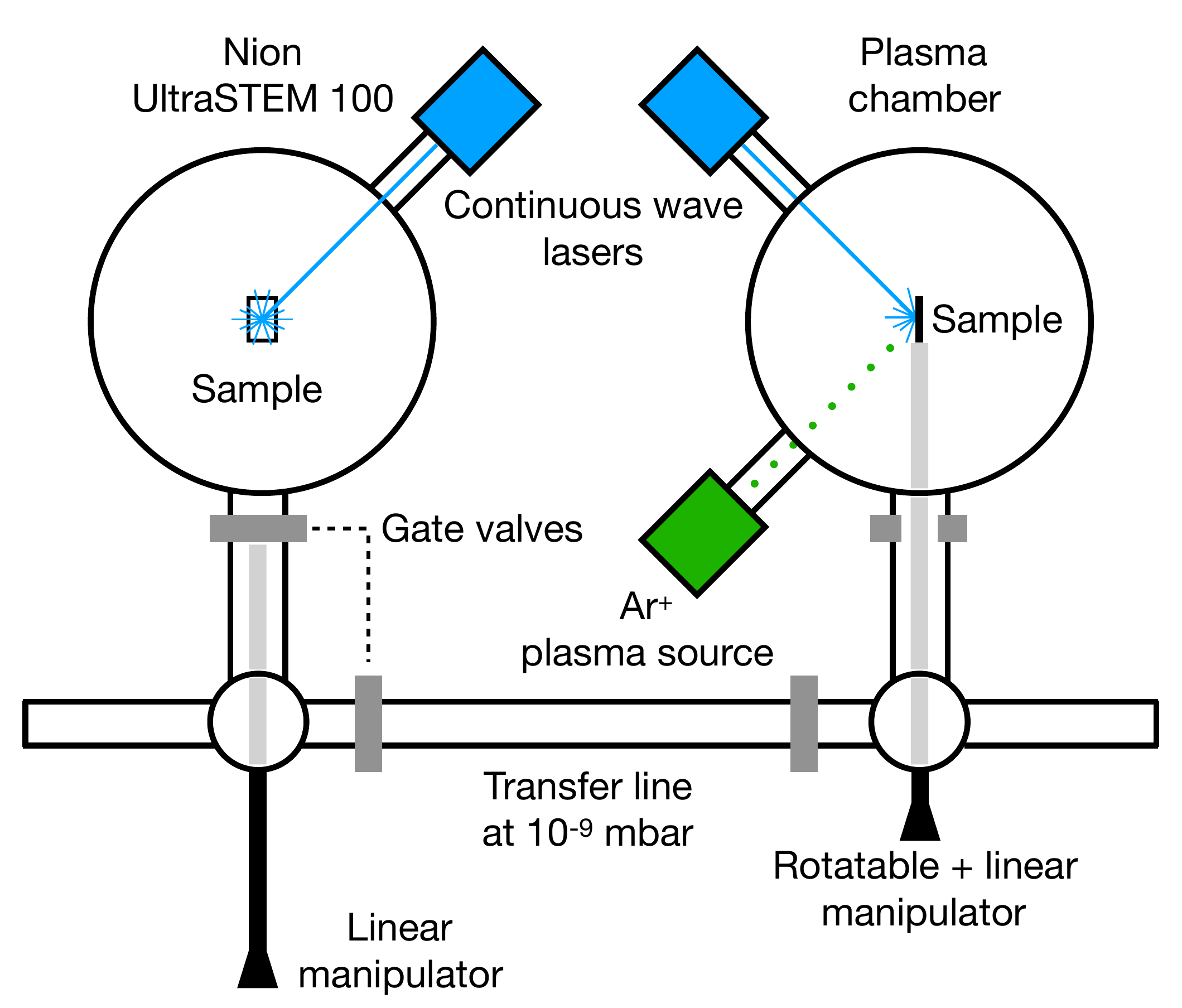}
\caption{\label{fig:figure_1} The experimental system consists of the modified aberration-corrected Nion UltraSTEM 100 scanning transmission electron microscope in Vienna~\cite{hotz_modified-nion-column_2016} connected to an external plasma chamber via an ultra-high vacuum transfer line. Both laser sources operate at 445~nm wavelength with a power tunable up to 6~W (Lasertack GmbH).~\cite{tripathi2017cleaning}}
\end{figure}

\begin{figure}
\includegraphics[width=8.6cm,keepaspectratio]{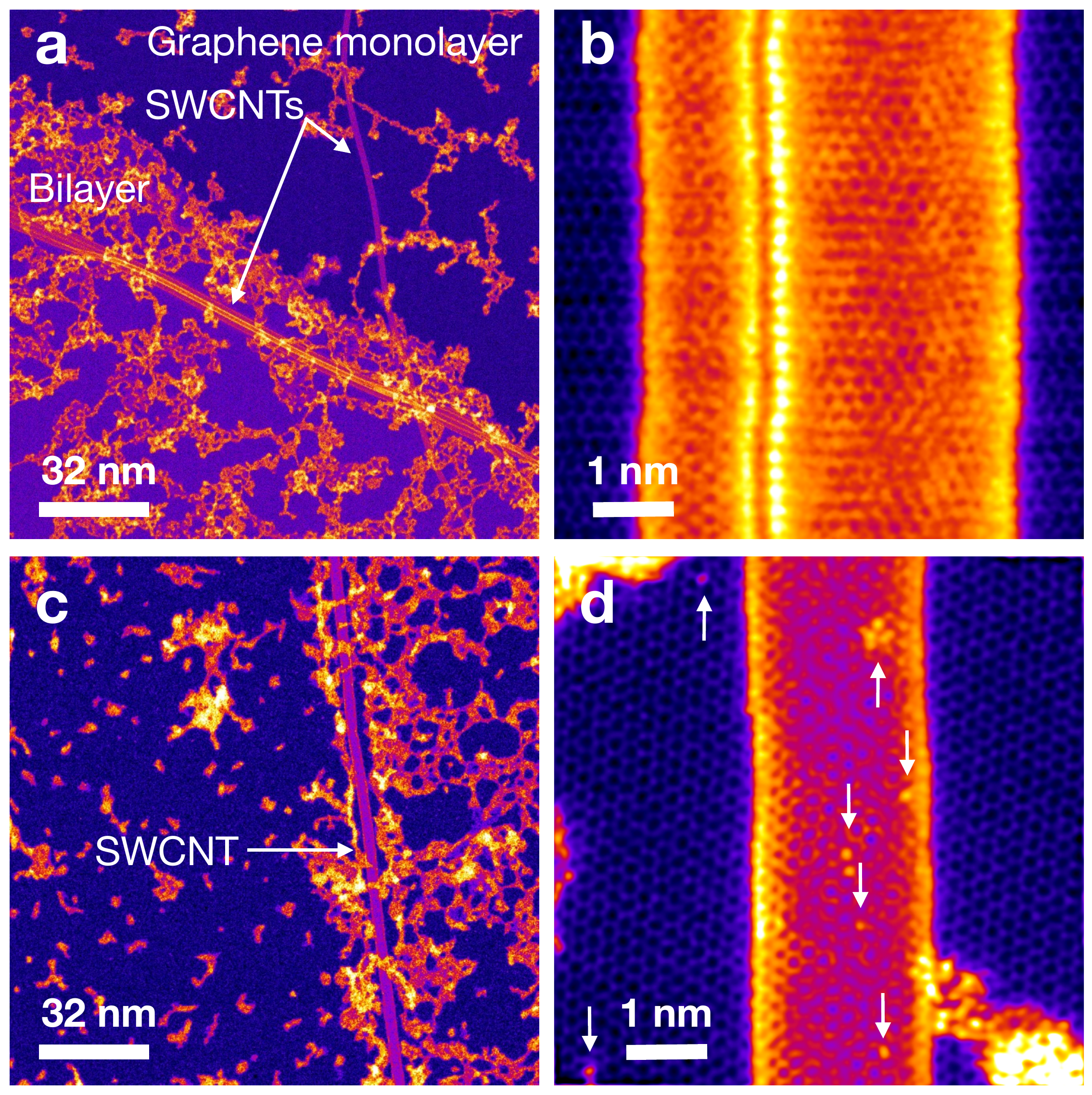}
\caption{\label{fig:figure_2} (a) STEM/MAADF overview of laser-cleaned SWCNTs on graphene. (b) An atomically resolved closeup of the interface. (c) An overview of a plasma-irradiated sample. (d) A closeup showing the presence of impurity atoms after plasma irradiation.}
\end{figure}

Motivated further by computational predictions of vacancy formation by argon (Ar) ions by Tolvanen \textit{et al.},~\cite{tolvanen2007relative} we demonstrate here the efficient covalent substitution of Si in both graphene and SWCNTs via an Ar plasma treatment. To allow a direct comparison, nanotubes were first grown in a floating catalyst reactor using ethanol and ferrocene as carbon and catalyst precursors~\cite{ding2017highly} and deposited on commercially available graphene on silicon nitride electron microscopy supports from Ted Pella Inc.~\cite{laiho2017dry, mustonen2018atomic} The deposition was followed by laser annealing in the Nion UltraSTEM 100 column,~\cite{tripathi2017cleaning} exposing clean surfaces for later plasma irradiation experiments (Figure~\ref{fig:figure_1}). An overview of the resulting clean tubes interfacing with graphene is shown in Figure~\ref{fig:figure_2}a and an atomically resolved closeup in Fig.~\ref{fig:figure_2}b. The images were acquired at 60~keV electron energy and a beam convergence semiangle of 30~mrad. The scattered electrons were detected with a medium angle annular dark field (MAADF) detector at an angular range of 60--200~mrad.

The cleaning was followed by plasma irradiation conducted in a purpose-built plasma target chamber directly connected to the Nion microscope via an ultra-high vacuum (UHV) transfer system (Fig.~\ref{fig:figure_1}). The argon plasma (pressure 5$\times$10$^{-6}$ mbar) was ignited in a microwave cavity on the side of the chamber and accelerated to a $\sim$50~eV kinetic energy, exposing the sample to a dose of $\sim$1~ion/nm$^{2}$. Chemical cross-linking and the buildup of amorphous contamination during the experiment was mitigated by concurrently applying $\sim$100 mW of laser power to the sample, which presumably simultaneously helped to mobilize the Si impurity atoms.~\cite{dyck2017placing} This resulted in the preservation of large clean areas, but also patched the contamination in small pockets as for example those visible in Fig.~\ref{fig:figure_2}c. Although not directly observable, these likely correspond to the positions of the greatest lattice disorder and therefore the highest chemical reactivity.

\begin{figure}
\includegraphics[width=8.6cm,keepaspectratio]{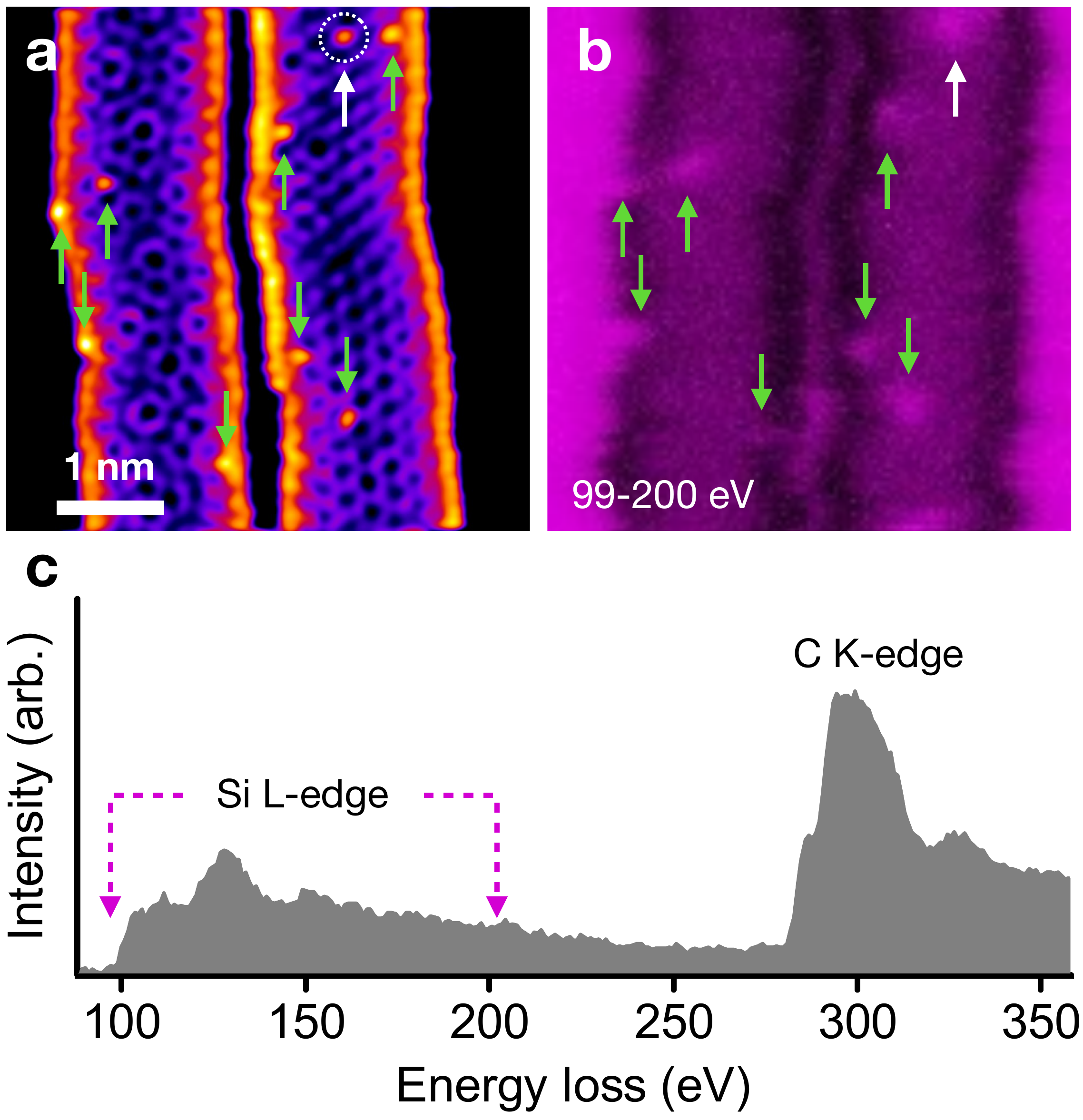}
\caption{\label{fig:figure_3} (a) An overview STEM/MAADF image of SWCNTs in vacuum and incorporating several covalent Si. (b) The mapped Si $L$-edge intensity in the 99--200 eV energy window (128~px $\times$ 128 px). (c) A spectrum acquired from the atom highlighted in (a) with spectral features consistent with 4-coordinated Si.~\cite{ramasse2013probing}}
\end{figure}

Even within the atomically clean areas and regardless of the laser irradiation, some vacancies especially in graphene remain unoccupied (Fig.~\ref{fig:SIfig1}). We nevertheless find a large number of covalently bound impurity atoms with a scattering contrast similar to Si~\cite{krivanek2010atom} (Fig.~\ref{fig:figure_2}d). To confirm their identity, we used electron energy loss spectroscopy (EELS) to acquire the elemental fingerprints of each atom.~\cite{ramasse2013probing,susi2017single} Since our focus was on impurities bound to SWCNTs, we studied the tubes suspended over the holes in the graphene support (Figure~\ref{fig:figure_3}a). The EELS system used here~\cite{susi2017single} consists of a Gatan PEELS~666 spectrometer with an energy dispersion of 0.5~eV/px and an Andor iXon 897 electron-multiplying camera. A background-subtracted spectrum recorded from the atom highlighted in Fig.~\ref{fig:figure_3}a is given in Fig.~\ref{fig:figure_3}c, indicating based on the $L$-edge shape and onset at $\sim$99~eV a 4-coordinated Si impurity.~\cite{ramasse2013probing,Zhou12PRL}. The spatial distribution of elements can be likewise analyzed by recording a spectrum from each probe position and by mapping the corresponding core losses. Using the Si $L$-edge intensity ($\sim$99--200~eV, see Fig.~\ref{fig:figure_3}c), the Si distribution was mapped and is visible in Fig.~\ref{fig:figure_3}b, confirming the chemical identity of the impurity atoms.

\begin{figure}
\includegraphics[width=8.5cm,keepaspectratio]{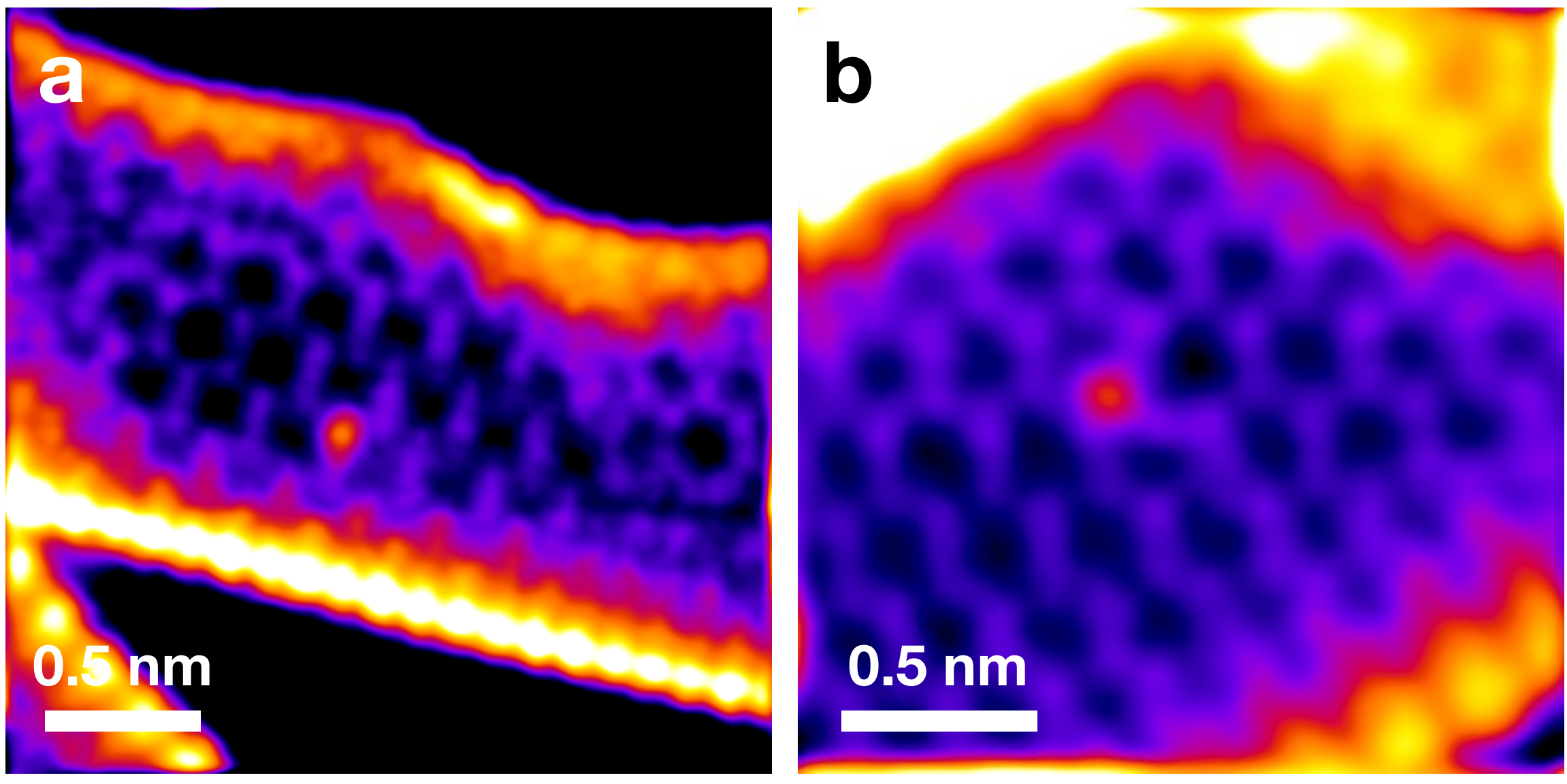}
\caption{\label{fig:figure_4} Examples of atomically resolved STEM/MAADF images of (a) Si-C$_3$ and (b) Si-C$_4$ impurities in SWCNTs.}
\end{figure}

We next turn our attention to the structure and abundance of the Si sites. The two fully saturated substitutions, 3- and 4-coordinated configurations (Si-C$_3$ and Si-C$_4$, respectively) have been identified in graphene.~\cite{ramasse2013probing,Zhou12PRL} Our atomic resolution observations, including those in Figure~\ref{fig:figure_4}, confirm that both configurations are also present in Ar-irradiated SWCNTs. We observed a total of 61~Si sites in 38~tubes (although the configuration could only be identified for 51), with $\sim$63\% being Si-C$_3$ and $\sim$37\% Si-C$_4$. These numbers agree remarkably well with the computationally projected abundance of single and double vacancies formed upon ion irradiation at energies similar to ours.~\cite{tolvanen2007relative} We likewise observed a large number of Si defects in graphene (Supplementary Figure~\ref{fig:SIfig1}), but their relative abundances were not determined.

Interestingly, regardless of the smaller energy required to displace C atoms from SWCNTs,~\cite{krasheninnikov_stability_2005} this was not reflected in the relative impurity density. We studied a total SWCNT surface area of 1200~nm$^2$ that contained the aforementioned 61~Si atoms, corresponding to an areal density of $\sim$0.05~nm$^{-2}$. A graphene surface area of 1365~nm$^2$ contained more than three times as many Si impurities, 210~atoms in total, giving a threefold areal density of $\sim$0.15~nm$^{-2}$. While Si impurities are commonly found on graphene~\cite{ramasse2013probing,Zhou12PRL,susi_silicon_2014} (although their origin remains unclear), they are not present on SWCNTs such as ours that have not undergone liquid dispersion. It is thus likely that to reach the vacancies created on nanotubes, the Si adatoms are first required to migrate over the graphene surface.

\begin{figure}
\includegraphics[width=8.5cm,keepaspectratio]{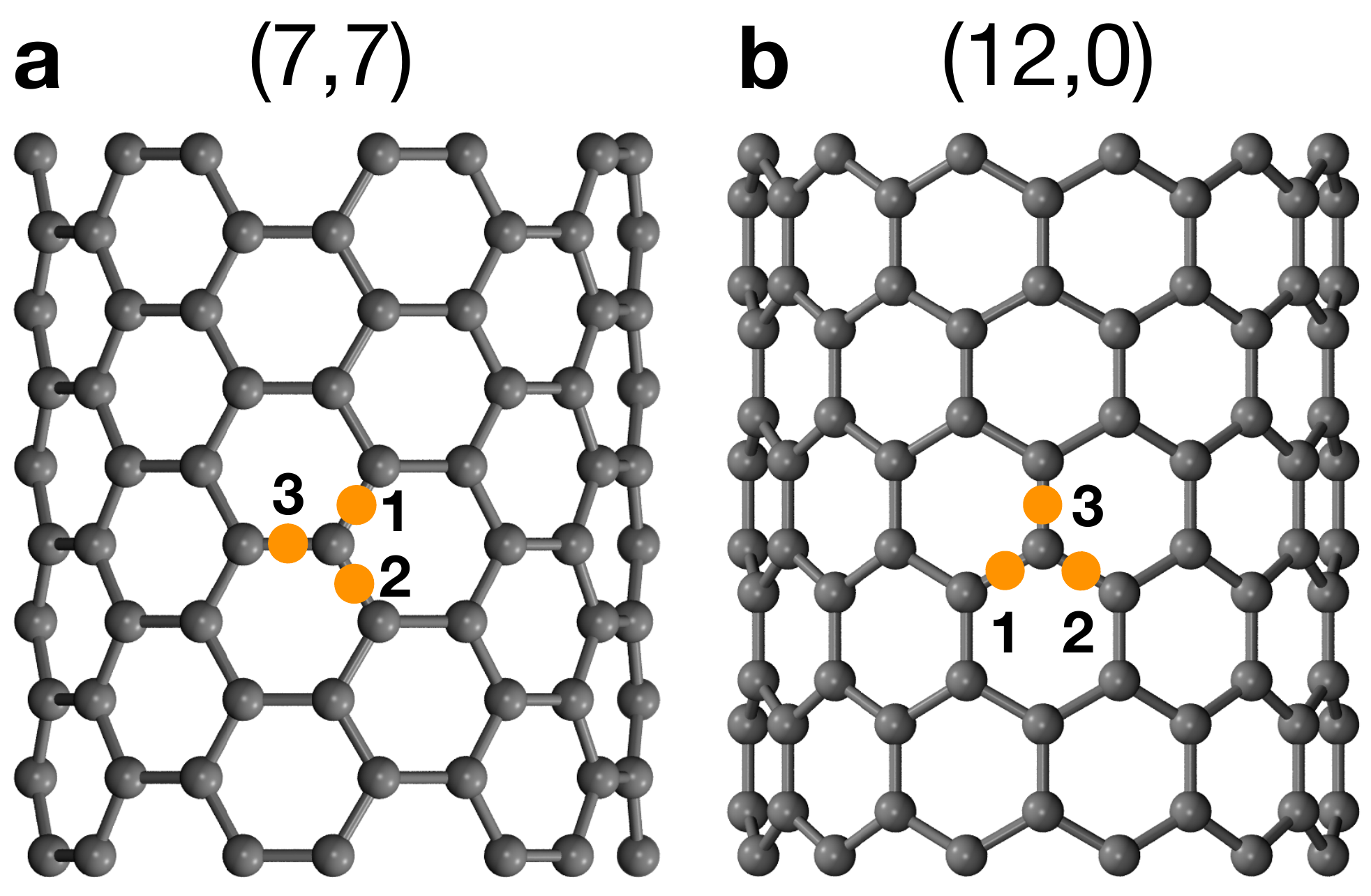}
\caption{\label{fig:figure_5} Si adsorption sites on (a) (7,7) armchair and (b) (12,0) zigzag SWCNTs reflecting two inequivalent migration paths (1-2 and 1-3).}
\end{figure}

To study the migration energetics, we ran density functional theory (DFT)-based atomistic simulations with the projector-augmented wave method implemented in the GPAW package,~\cite{enkovaara_electronic_2010} and calculated the binding energies ($E_\text{b}$) and migration barriers ($E_\text{m}$) of Si adatoms on graphene and on a set of achiral SWCNTs (Table~\ref{tab:table1} and Figure~\ref{fig:figure_5}). We used the revPBE~\cite{hammer_improved_1999} exchange-correlation functional, a \textit{dzp} basis set,~\cite{larsen_localized_2009} and a grid spacing of 0.2~\AA. The length of the models for armchair and zigzag \mbox{SWCNTs} were 12.35 and 12.83~\AA, respectively. Graphene was modelled using a 7$\times$7 supercell of 98 atoms. The Brillouin zone was sampled using 8 \textbf{k}-points in the periodic axial direction for nanotubes and a 6$\times$6$\times$1 \textbf{k}-point grid for graphene. Migration barriers were obtained with the climbing-image nudged elastic band method (cNEB).~\cite{henkelman_climbing_2000} 

The lowest energy configurations for Si adatoms on both graphene and nanotubes correspond to the bridge site, i.e. above a C-C bond. The calculated binding energy on graphene was 0.34~eV, slightly lower than the previously reported values (0.44 eV~\cite{xian2012diffusion} and 0.55 eV~\cite{pavsti2018atomic}). This discrepancy appears to arise from differences in exchange-correlation functionals, revPBE vs. PBE~\cite{perdew_generalized_1996}, since with the latter $E_\text{b}$ = 0.55 eV was also reproduced in our calculations. On the surface of nanotubes, the binding energies depend on the tube diameter, chirality and adsorption site. For smaller diameter tubes the calculated $E_\text{b}$ values are significantly higher than those for graphene, while converging to the graphene value as the tube diameter increases. Interestingly, the most reactive C-C bridge never lies on the tubes' axis or along their circumference, but is instead diagonal (sites 1 and 2 in Fig.~\ref{fig:figure_5}). This is different from C adatoms that have been shown to prefer the circumferential bridge configuration on armchair tubes.~\cite{krasheninnikov2004adsorption}

The calculated energy barriers (Table~\ref{tab:table1}) show that migration on armchair SWCNTs prefers the direction of the nanotube axis. For the (7,7) and (15,15) tubes, migration barriers along the path 1-2 are only 0.08 eV and 0.06 eV, respectively, and are therefore very close to the value in graphene (0.06~eV in our calculations, in excellent agreement with previously reported values of 0.06--0.07~eV).~\cite{xian2012diffusion} By contrast, migration paths around the circumference have barriers 3--5~times higher and are therefore much less likely to occur. Migration on zigzag tubes is less directed. According to cNEB calculations for the (12,0) tube, migration along the path 1-2 on the tube circumference always occurs via configuration 3. Meanwhile, the Si adatom jumps between positions that are equivalent to sites 1 and 3 can occur in several directions. The calculated energy barriers for these jumps are higher than those for migration along the axis of armchair nanotubes and therefore migration of Si adatoms on zigzag tubes will be slower.

It is important to note that the Si migration barriers are much lower than those of C adatoms (0.47~eV on graphene~\cite{lehtinen2003magnetic} and  $\sim$0.6--1.3~eV on 10--15~\AA\ diameter nanotubes~\cite{krasheninnikov2004adsorption}), which may explain why Si is such a common lattice impurity.~\cite{ramasse2013probing,susi_silicon_2014}. Since the barriers in larger diameter tubes are further nearly as low as in graphene, differences in Si migration over the surface of the two materials do not provide a direct explanation for their differing Si concentrations. The fact that we observed many fewer residual vacancies in SWCNTs, however, suggests that C reconstruction in nanotubes is in general more efficient. One possible explanation is that endohedral C adatoms (either endemic or sputtered by the Ar ions) have migration barriers much smaller than those on the outer wall.~\cite{krasheninnikov2004adsorption} Since these atoms are unable to escape from the quasi-one-dimensional tube, each migrating C will rapidly sample the entire inner surface and recombine with any vacancy it encounters,~\cite{gan_diffusion_2008} effectively hindering the incorporation of Si by reducing the number of available substitution sites.

\begin{table}
\caption{\label{tab:table1} Calculated values of binding energies ($E_\text{b}$) and migration barriers ($E_\text{m}$) for Si adatoms on graphene and SWNTs. The corresponding adsorption sites are shown in Fig.~\ref{fig:figure_5}.}
\begin{ruledtabular}
\begin{tabular}{ccccc}
CNT & Site & $E_\text{b}$, eV & Path & $E_\text{m}$, eV\\
\hline
 Graphene &  & 0.34 &  & 0.06\\
 \\
(7,7) & 1 & 0.82 & 1-2 & 0.08\\
(7,7) & 3 & 0.66 & 1-3 & 0.43\\
(7,7) &  &       & 3-1 & 0.27\\
\\
(15,15) & 1 & 0.46 & 1-2 & 0.06\\
(15,15) & 3 & 0.32 & 1-3 & 0.26\\
(15,15) &   &      & 3-1 & 0.12\\
\\
(12,0) & 1 & 0.95 & 1-3 & 0.35\\
(12,0) & 3 & 0.74 & 3-1 & 0.13\\
\\
(26,0) & 1 & 0.43 & 1-3 & 0.19\\
(26,0) & 3 & 0.33 & 3-1 & 0.09\\
\end{tabular}
\end{ruledtabular}
\end{table}

To summarize, we have used argon ions to create mono- and divacancies in graphene and single-walled carbon nanotubes and demonstrated the substitution of Si impurities in both materials in respective concentrations of 0.15~nm$^{-2}$ and 0.05~nm$^{-2}$. The captured Si bonded in both 3- and 4-coordinated configurations, directly identified using atomically resolved scanning transmission electron microscopy and electron energy loss spectroscopy. Finally, our atomistic simulations show that Si migration is substantially faster than that of C. Our findings could enable also other impurities with similar migration barriers to be captured in the graphitic lattice in greater quantities than has so far been achieved.

\begin{acknowledgments}
M.T. and T.S. acknowledge the Austrian Science Fund (FWF) project P 28322-N36 for funding, and T.S. and A.M. also the European Research Council (ERC) Grant No. 756277-ATMEN. E.I.K, E.D. and A.H. acknowledge the support from the Academy of Finland \textit{via} projects 286546-DEMEC and 292600-SUPER, from TEKES Finland \textit{via} projects 3303/31/2015 (CNT-PV) and 1882/31/2016 (FEDOC), and the Aalto Energy Efficiency (AEF) Research Program through the MOPPI project. A.M. and T.S. acknowledge the Vienna Scientific Cluster for computer time. J.K., H.I. and K.M. were supported by the FWF project I3181 and the Wiener Wissenschafts\mbox{-,} Forschungs- und Technologiefonds (WWTF) project MA14-009, and K.M. further by the Finnish Cultural Foundation through a grant from the Finnish Postdoc Pool and J.K. through FWF project P31605.
\end{acknowledgments}

\bibliographystyle{apsrev4-1}
\bibliography{aipsamp}

\pagestyle{empty}
\clearpage
\onecolumngrid
\section*{Supporting Information}

\renewcommand{\thefigure}{S\arabic{figure}}
\setcounter{figure}{0}

\begin{figure*}[h]
\includegraphics[width=15.5cm,keepaspectratio]{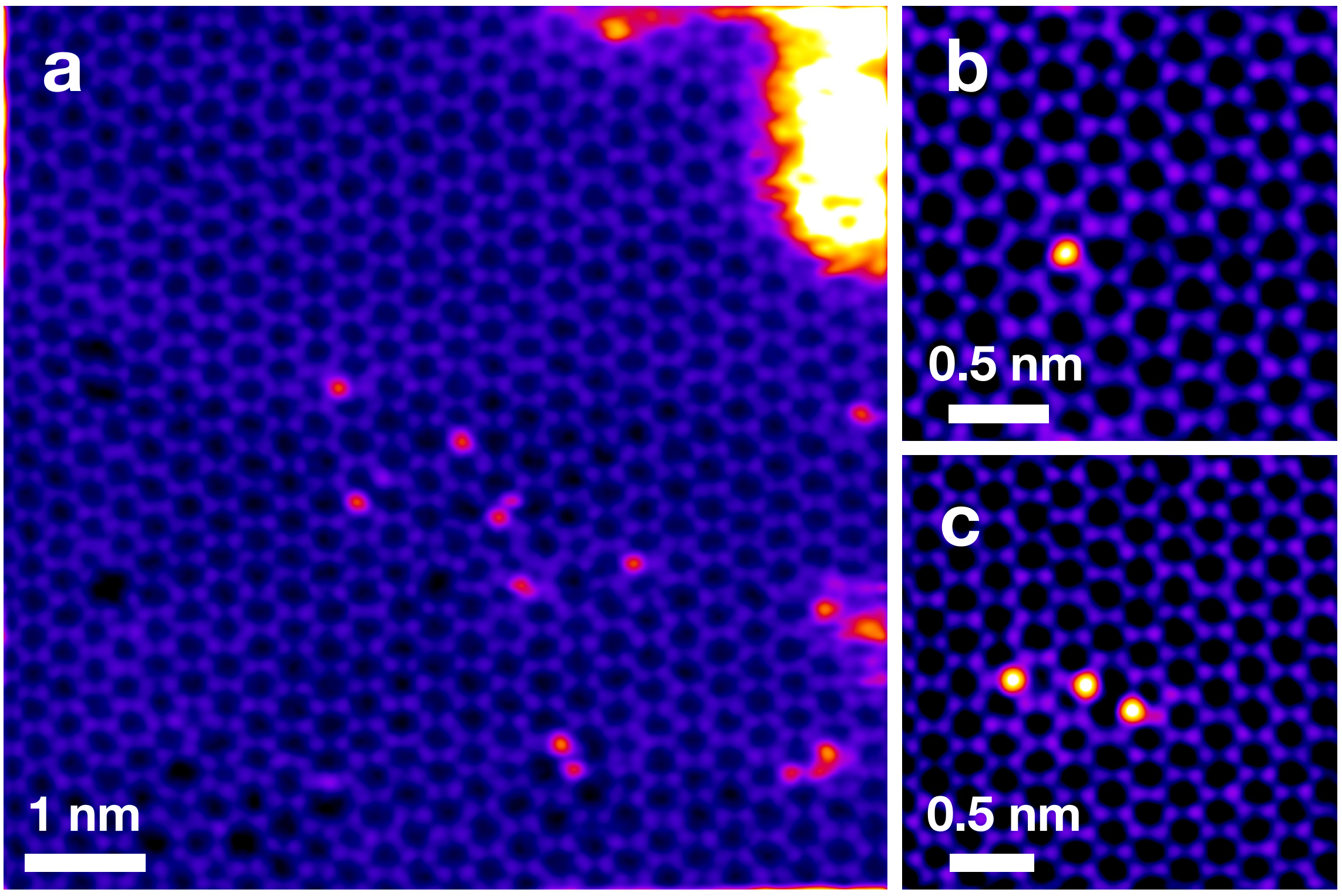}
\caption{STEM/MAADF images of silicon-substituted graphene. a) Overview of the lattice with many point defects as well as individual and clustered impurities. b) An individual silicon atom in 4-coordinated configuration. c) Three nearby silicon impurities.\label{fig:SIfig1}}
\end{figure*}

\end{document}